\documentclass{elsart3p}

\usepackage{graphicx}

\usepackage{amssymb}

\begin{document}

\begin{frontmatter}



\title{Muon Spin Rotation and the Vortex Lattice in Superconductors}


\author[EHB]{Ernst Helmut Brandt\corauthref{cor}}
\corauth[cor]{email: ehb@mf.mpg.de}

\address[EHB]{Max Planck Institute for Metals Research, D-70506 Stuttgart, Germany}

\begin{abstract}
The magnetic field probability $P(B)$ is calculated
from Ginzburg-Landau theory for various lattices of vortex lines
in type-II superconductors: Ideal triangular lattices, lattices
with various shear strains and with a super lattice of
vacancies, and lattices of short vortices in films whose
magnetic field ``mushrooms'' near the surface.
\end{abstract}

\end{frontmatter}

\section{Introduction}
\label{sec:introduction}

Type-II superconductors like Niobium and many alloys allow magnetic
flux to penetrate in form of magnetic flux lines, i.e., vortices
of the supercurrent, each vortex carrying one quantum of magnetic
flux. This effect was predicted in 1957 by Alexei Abrikosov
\cite{1}, who got for this the Nobel Prize in Physics 2003.
Abrikosov flux lines arrange to a more or less perfect triangular
vortex lattice that exhibits interesting structural defects that
may be calculated from Ginzburg-Landau (GL) theory \cite{2} or by
treating the vortex lattice as a continuum with non-local
elasticity \cite{3}. The vortex lattice may even melt into a
``vortex liquid'' \cite{4}. Pinning of vortices by material
inhomogeneities \cite{5} together with thermal fluctuations of the
vortices can cause a rich phase diagram in the magnetic
field--temperature plane \cite{6}, with a melting line and an
order--disorder line\cite{7,8} at which the weak elastic disorder
(``Bragg glass'') suddenly changes to a plastically deformed or
even amorphous vortex arrangement. The vortex lattice can be
observed by decoration, magneto-optics, Hall probes, neutron
scattering, magnetic force microscopy, and by muon spin rotation
($\mu$SR). $\mu$SR experiments can give valuable information
about the vortex lattice, see the recent review \cite{9}.

  When pinning and thermal fluctuations may be disregarded (e.g.,
in clean Niobium with very weak pinning) the vortex lattice exists
when the applied magnetic field $B_a$ lies between the lower
critical field $B_{c1}$ and the upper critical field $B_{c2}$. In
this interval the internal average induction  $\bar B$ is smaller
than $B_a$, i.e., the magnetization $M = \bar B -B_a$ is negative
(diamagnetic behavior), ranging from $-B_{c1}$ to 0 while $\bar B$
ranges from 0 to $B_{c2}$. For $B_a < B_{c1}$ the superconductor
expels the magnetic field ($\bar B =0$, ideal Meissner state) and
for $B_a > B_{c2}$ the superconductor is in the normal conducting
state ($\bar B = B_a$). This applies to long superconductor
cylinders or slabs in parallel $B_a$. For other geometries and for
inhomogeneous materials, demagnetization effects modify this
picture and the magnetization curve in general has to be computed
numerically, e.g., for thick or thin strips, disks, and platelets
in perpendicular $B_a$ \cite{10}. However, in the particular case
when a homogeneous specimen with the shape of an ellipsoid is put
into a uniform applied field ${\bf B}_a$, then the demagnetizing
field (caused be the magnetization) inside the ellipsoid is also
uniform and superimposes to ${\bf B}_a$, thus generating an
effective applied field
  \begin{eqnarray}  
  {\bf B}_i = {\bf B}_a -N {\bf M(B}_i; N=0).
  \end{eqnarray}
Solving Eq.~(1) for the effective internal field ${\bf B}_i$, one
obtains ${\bf M}={\bf M(B}_a; N) ={\bf M(B}_i ;N=0)$. Here $N$ is
the demagnetization factor, and the ideal magnetization curve
[$M(B_a; 0)$ for $N=0$, e.g., from GL theory] should be inserted.
In general, $N$ is a tensor, but when $B_a$ is along one of the
three principal axes of the ellipsoid, then $N$ and all fields in
(1) are scalars. For long cylinders or slabs in parallel $B_a$ one
has $N=0$, for spheres $N=1/3$, for long cylinders in
perpendicular field $N=1/2$, and for thin plates and films one has
$1-N \ll 1$. The sum of the $N$ along the three axes is always
$N_1+N_2+N_3 = 1$. For the Meissner state one finds $\bar B =0$,
$M(B_a; 0) = -B_a$, $B_i =B_a/(1-N)$, and $M(B_a;N) =-B_a/(1-N)$,
which means the vortex penetration starts at the effective
penetration field $B_{c1}' = (1-N) B_{c1}$ where $B_i = B_{c1}$.
In the interval $B_{c1}' < B_a < B_{c2}$ such a pin-free ellipsoid
contains a perfect vortex lattice.

In this paper the probability $P(B)$ that at a random point inside
the superconductor a muon sees an induction value $B$, is
considered for ideal triangular vortex lattices and for various
possible perturbations of  it, namely, various types of shear
deformation \cite{11}, a super lattice of vortex vacancies, the
surfaces of a film in a perpendicular field, and random
displacements. This probability (or field density) is defined as
the spatial average
  \begin{eqnarray}  
  P(B') = \big\langle \delta \big(B' -
          B({\bf r}) \big) \big\rangle_{\bf r} .
  \end{eqnarray}
In it $B'$ is the independent variable, $B({\bf r})$ the spatially
varying magnetic field, and $\delta(x)$ is the 1D delta function,
which for computation may be replaced by a narrow Gaussian whose
width may depend on $B'$. One easily shows that
  \begin{eqnarray}  
   \int\nolimits_{-\infty}^{\,\infty} \!P(B)\,dB = 1, ~~~
   \int\nolimits_{-\infty}^{\,\infty} \!P(B)\, B\,dB = \bar B,
  \end{eqnarray}

\begin{figure}[htb]  
\centering
\includegraphics[width=\columnwidth]{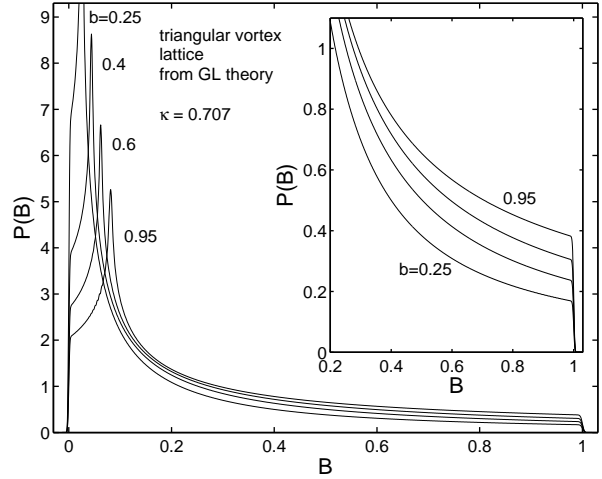}
\vspace{-2mm}
\caption{The field density $P(B)$ of the ideal triangular
vortex lattice obtained from GL theory for
$\kappa = 1/\sqrt2 =0.707$ and $b = 0.25$ to 0.95.
Here and in the figures below
the variable $B$ is normalized such that $B=0$ means
$B_{\rm min}$ and $B=1$ means $B_{\rm max}$, and
the inset enlarges the region near $B_{\rm max}$.}
\end{figure}

\begin{figure}[htb]  
\centering
\includegraphics[width=\columnwidth]{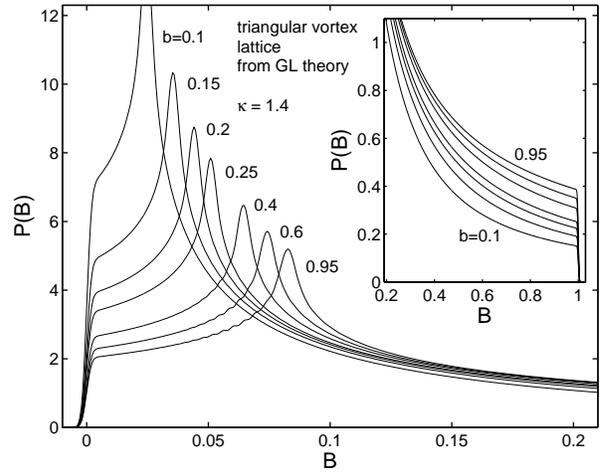}
\vspace{-2mm}
\caption{The $P(B)$ as in Fig.~1 but for for $\kappa=1.4$.}
\end{figure}

\begin{figure}[htb]  
\centering
\includegraphics[width=\columnwidth]{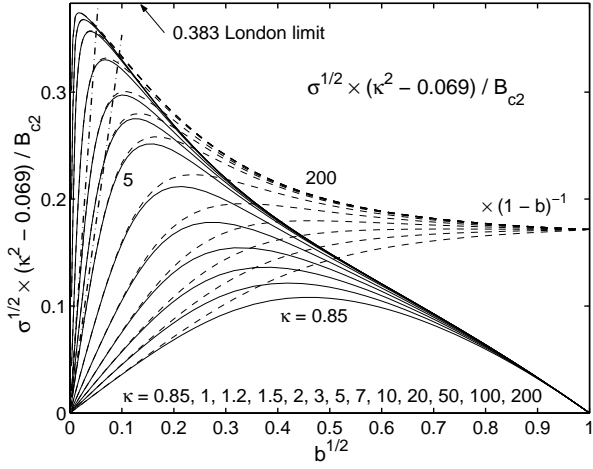}
\vspace{-2mm}
\caption{The magnetic field variance $\sigma = \langle [B(x,y) -
\bar B]^2 \rangle$ of the triangular FLL for $\kappa=0.85$ to
200 plotted in units of $B_{c2}$ as
$\sqrt{\sigma}\cdot(\kappa^2-0.069)/B_{c2}$ (solid lines) such
that the curves for all $\kappa$ collapse near $b=1$. The dashed
lines show the same functions divided by $(1-b)$ such that they
tend to a finite constant 0.172 at $b=1$. All curves are plotted
versus $\sqrt b = \sqrt{\bar B/B_{c2}}$ to stretch them at small $b$
values and show that they go to zero linearly. The upper frame
0.383 is the usual London approximation. The limit for very small
$b$ is shown as two dash-dotted straight lines for $\kappa=5$ and
$\kappa =10$. The upper frame 0.383 shows the usual London
approximation.}
\end{figure}

\section{2D ideal triangular vortex lattice }

   The local magnetic field $B(x,y)$ of the vortex lattice may
be calculated for all values of the reduced average induction
$b = \bar B/B_{c2}$ and Ginzburg-Landau parameter $\kappa$ by
an elegant iteration method that minimizes the GL free energy
$F$ with respect to the Fourier coefficients of the periodic
solutions $B(x,y)$ and order parameter
$\omega(x,y) =f^2=|\psi|^2$ where $\psi(x,y)=f\exp(i\varphi)$
is the complex GL function \cite{12,13}.
In the usual reduced units (length $\lambda$, induction
$\sqrt2 B_c$, energy density $B_c^2/\mu_0$, where $\lambda$
is the magnetic penetration depth and
$B_c = B_{c2}/\sqrt2 \kappa$ is the thermodynamic critical
field with $B_{c2}= \Phi_0 /(2\pi \xi^2)$, $\xi=\lambda/\kappa$
the coherence length and
$\Phi_0 = h/2e = 2.07\cdot 10^{-15}$ Tm$^2$ the quantum of
magnetic flux) the spatially averaged free
energy density $F$ of the GL theory referred to the Meissner
state ($\psi=1$, ${\bf B} =0$) in the superconductor reads
\begin{equation}  
  F = \left\langle {(1-|\psi|^2)^2 \over 2} + \left| \left(
  {\nabla \over i\kappa} - {\bf A} \right) \psi \right|^2
  + {\bf B}^2 \right\rangle .
  \end{equation}
Here ${\bf B(r)}=\nabla\times {\bf A}$,
${\bf A(r)}$ is the vector potential, and
$\langle \dots \rangle = (1/V) \int_V d^3r \dots $ means
spatial averaging over the superconductor with volume $V$.
Introducing the super velocity
${\bf Q}({\bf r}) = {\bf A} -\nabla\varphi/\kappa$ and the
magnitude $f({\bf r}) = |\psi|$ one may write $F$ as a
functional of these real and gauge-invariant
functions, 
  \begin{equation}  
  F = \left\langle {(1-f^2)^2 \over 2} + {(\nabla f)^2 \over
  \kappa^2 } +f^2 Q^2 +(\nabla\! \times\! {\bf Q})^2 \right\rangle.
  \end{equation}
In the presence of vortices ${\bf Q}({\bf r})$ has to be chosen
such that $\nabla\! \times\! {\bf Q}$ has the appropriate
singularities along the vortex cores, where $f$ vanishes.
By minimizing this $F$ with respect to $\psi$, ${\bf A}$ or
$f$, ${\bf Q}$, one obtains the two GL equations together with
the appropriate boundary conditions. For periodic lattices
with one flux quantum per vortex, in the sense of a Ritz
variational method one uses Fourier series for the periodic
trial functions with a finite number of Fourier coefficients
$a_{\bf K}$ and $b_{\bf K}$,
  \begin{eqnarray}  
  \omega({\bf r})&=&\sum_{\bf K} a_{\bf K} (1-\cos{\bf K r})\,,\\
  B({\bf r}) &=& \bar B +\sum_{{\bf K}\ne 0} b_{\bf K}\,
                 \cos{\bf K r}\,,\\
  {\bf Q(r)} &=& {\bf Q}_A({\bf r}) + \sum_{{\bf K}\ne0} b_{\bf K}
  {{\bf\hat z \times K} \over K^2} \sin{\bf K r} \,,
  \end{eqnarray}
where ${\bf K = K}_{mn} = (K_x,K_y)$
are the reciprocal lattice vectors of the vortex
lattice with positions ${\bf R}_{mn}$,
\begin{eqnarray}  
    {\bf R}_{mn} &=&(mx_1 +nx_2;\, ny_2) \,,\\
    {\bf K}_{mn} &=&(2\pi/x_1y_2)(my_2;\, -mx_2 +nx_1) \,,
  \end{eqnarray}
($m,n = 0, \pm1, \pm2, \dots$; triangular lattice: $x_1=a$,
$x_2=x_1/2$, $y_2=x_1\sqrt3/2$; square lattice: $x_1=y_2=a$,
$x_2=0$). In (8) ${\bf Q}_A(x,y)$ is the supervelocity of the
Abrikosov $B_{c2}$ solution, which satisfies
  \begin{eqnarray}    
  \nabla \times {\bf Q}_A = \Big[ \bar B -\Phi_0 \sum_{\bf R}
  \delta_2({\bf r-R}) \Big] {\bf\hat z}\,,
  \end{eqnarray}
where $\delta_2({\bf r}) = \delta(x)\delta(y)$ is the 2D delta
function. This shows that ${\bf Q}_A$ is the velocity field of a
lattice of ideal vortex lines but with zero average rotation.
Close to each vortex center one has ${\bf Q}_A({\bf r}) \approx
{\bf r' \times\hat z}/(2\kappa r'^2)$ and $\omega({\bf r}) \propto
r'^2$ with ${\bf r'= r-R}$. We take ${\bf Q}_A$ from
the exact relation
  \begin{eqnarray}    
  {\bf Q}_A({\bf r}) = {\nabla \omega_A \times {\bf\hat z}
  \over 2\, \kappa\, \omega_A } \,,
  \end{eqnarray}
where $\omega_A(x,y)$ is the Abrikosov $B_{c2}$ solution given
by a rapidly converging series (6) with coefficients
$a_{\bf K}^A = -(-)^{mn+m+n} \exp(-K_{mn}^2 x_1y_2 / 8\pi)$.
The solutions $\omega({\bf r})$ and $B({\bf r})$ are then
computed by iterating equations for the coefficients
$a_{\bf K}$ and $b_{\bf K}$ that are derived from the GL equations
$\delta F /\delta \omega =0$ and $\delta F /\delta {\bf Q} =0$.
This method and the obtained GL solutions are presented
in detail in \cite{12,13}.

  The field density $P(B)$ of the ideal triangular vortex
lattice is shown for several $b =\bar B/B_{c2}$ values in Fig.~1
($\kappa = 1/\sqrt2$) and Fig.~2 ($\kappa=1.4$). For larger
$\kappa$ the $P(B)$ look similar to Fig.~2. For small $b$ and
large $\kappa$ the $P(B)$ obtained from the London approximation
are depicted in \cite{11}. The maximum and the 2 equal minima
per unit cell yield two steps in $P(B)$ at $B=B_{\rm max}$ (=1)
and  $B=B_{\rm min}$ (=0), and the 3 equal saddle points yield
a logarithmic infinity  at $B=B_{\rm sad}$ where
$P(B) \propto -\ln |B -B_{\rm sad}|$.

The variance of the magnetic field,
$\sigma =\langle [B(x,y) -\bar B]^2 \rangle =\sum_{{\bf K}\ne 0}
b_{\bf K}^2$, is plotted for the entire ranges of $b$ and
$\kappa$ in Fig.~3. In the low-field range
$0.13/\kappa^2 \ll b \ll 1$ one has for the triangular lattice
the London limit $\sigma =0.00371 \Phi_0^2/\lambda^4 $
(upper frame in Fig.~3), at very small $b \ll 0.13/\kappa^2$
one has $\sigma =(b \kappa^2 /8\pi^2) \Phi_0^2/\lambda^4 $
(dash-dotted straight lines in Fig.~3), and near $b=1$
one has the Abrikosov limit $\sigma =7.52\cdot 10^{-4}
(\Phi_0^2/\lambda^4) [\kappa^2 (1-b) / (\kappa^2 -0.069)]^2$,
approximately valid even at $b \gtrapprox 0.3$, see \cite{13}.
Note that the usual London limit for $\sigma$
applies only in a narrow range of small (but not too small)
$b$ and for $\kappa > 50$. The same is true for the London
limit of the magnetization curve, where the often used
``logarithmic law valid at $B_{c1} \ll B_a \ll B_{c2}$''
for $M(B_a) =\bar B -B_a$ has a small range of
validity \cite{13}.

\begin{figure}[htb]  
\centering
\includegraphics[width=\columnwidth]{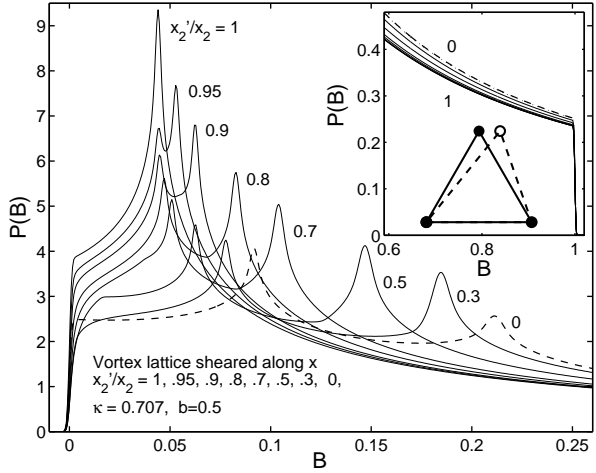}
\vspace{-2mm}
\caption{Vortex lattice sheared along the $x$ axis.}
\end{figure}

\begin{figure}[htb]  
\centering
\includegraphics[width=\columnwidth]{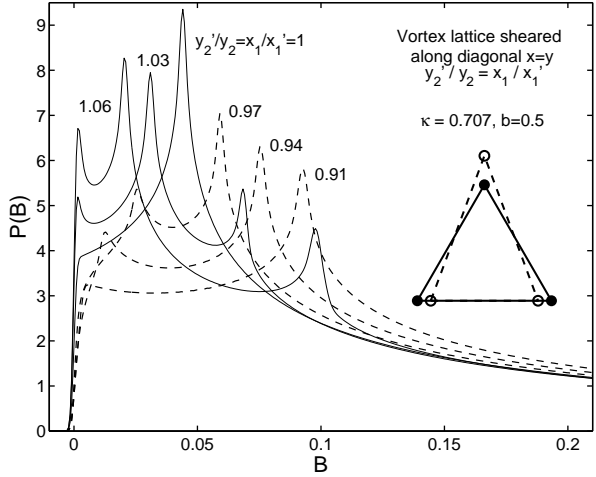}
\vspace{-2mm}
\caption{Vortex lattice sheared along the diagonal $x=y$.}
\end{figure}

\section{Sheared vortex lattices}

  The above Fourier method applies to any vortex lattice symmetry
with vortex positions ${\bf R}_{mn}$ (9),  also to sheared
triangular lattices and to square and rectangular basic cells.
In Fig.~4 the field density $P(B)$ is shown for a lattice sheared
away from the ideal triangular lattice by decreasing in Eq.~(9)
the length $x_2 =0.5$ by a factor $c= x_2' / x_2 = 1$ (triangular),
0.95, 0.9, $\dots$, 0 (rectangular lattice). With $y_2=\sqrt3 /2$
the shear strain in these cases is $\gamma = (1-c)\,x_2/y_2 =
(1-c)/\sqrt3$. Note that for this shear the saddle point peak
splits into 2 peaks, i.e., there are now two different types of
saddle points in $B(x,y)$, but still one maximum and 2 equal minima.
Figure~5 shows a different orientation of shearing the triangular
lattice, namely, now the lengths $y_2$ and $x_1$ are changed by
a factor $c = y_2'/y_2 = x_1/x_1'$ = 1.06, 1.03, 1, 0.97, 0.94,
0.91 such that the unit cell area $x_1 y_2 = x_1' y_2'$ does not
change (no compression). This corresponds to a shear strain
of size $\gamma =2(1-c)$ oriented along the diagonal $x=y$.
One can see that for $c > 1$ the saddle point peak splits into
3 peaks (i.e. all 3 saddle points now occur at different $B$)
while for $c<1$ there occur 2 different saddle point peaks.

   One notes that even very small shear of the vortex lattice
causes pronounced change in the field probability $P(B)$. Small
shear costs very little energy since the shear modulus $c_{66}$
of the vortex lattice is much smaller than its compressional
modulus $c_{11}$ or its tilt modulus $c_{44}$. One has
approximately \cite{3}
 \begin{eqnarray}  
  c_{11}(k) = {\bar B^2 \over \mu_0} {\partial B_a \over\partial
    \bar B }\, {1 \over (1 +k^2 \lambda'^2) }
    + c_{66} \\
  c_{66} \approx {\bar B B_{c2} \over 8 \kappa^2 \mu_0}\,
     {(1-b)^2 (2\kappa^2 -1) 2\kappa^2  \over
    ( 2\kappa^2 -1 +1/\beta_A )^2 }  \\
  c_{44}(k) = {\bar B^2 \over \mu_0}\, {1 \over1+k^2\lambda'^2}
    + { \bar B (B_a -\bar B) \over \mu_0}
   \end{eqnarray}
with $\lambda' =\lambda/ \sqrt{1-b}$.
In $c_{66}$, $\beta_A = 1.1596$ is the Abrikosov parameter
of the triangular lattice (the square lattice is unstable
and thus has negative $c_{66}$), and the factor $(2\kappa^2-1)$
means the shear stiffness of the vortex lattice is zero
in superconductors with $\kappa = 0.707$ (pure Nb).

  An interesting property is the dependence
of $c_{11}$ (13) and $c_{44}$ (15) on the magnitude
$k = |{\bf k}|$ of the wave vector ${\bf k}=(k_x,k_y,k_z)$
of spatially periodic strain,
which means the elasticity of the vortex lattice is
{\bf non-local}. In the limit of uniform stress,
$k \to 0$, these expressions reproduce the known values
of the uniform compression and tilt
moduli obtained by thermodynamics, $c_{11} -c_{66}
 = (\bar B^2 / \mu_0) \partial Ba / \partial \bar B$,
$c_{44} =\bar B B_a / \mu_0$. However, when the wavelength
of the periodic compression or tilt decreases, i.e.,
$k$ increases, these moduli decrease. This
means, the vortex lattice is softer for short-wavelengths
compression and tilt than it is for long wavelengths.
In anisotropic superconductors these moduli at finite
wavelengths are even smaller \cite{14} and the vortex
lattice is softer and can be distorted and melted more
easily in high-$T_c$ superconductors.

\begin{figure}[htb]  
\centering
\includegraphics[width=\columnwidth]{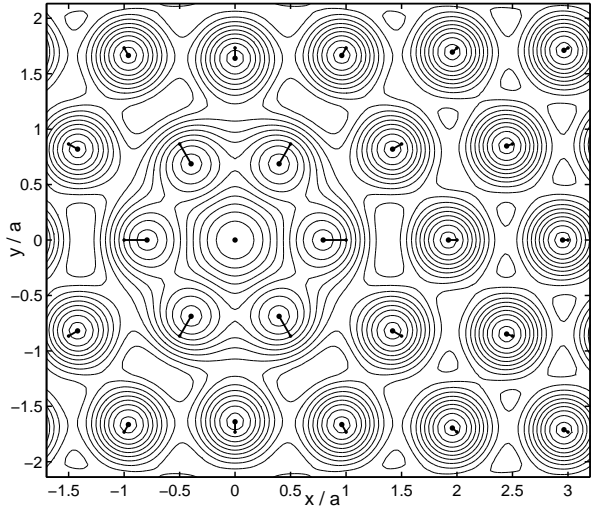}
\caption{Contour lines of order parameter $\omega(x,y)$,
Eq.~(17), identical to the contours of induction  $B(x,y)$,
Eq.~(16), for a vortex lattice with one vacancy. The
displacements ${\bf s}_\nu$ of the relaxing vortices are
shown as short bold lines connecting two dots. At all
vortex positions $\omega(x,y)$ has a minimum and is zero,
but at the origin $x=y=0$, $\omega$ is maximum since a
vortex was removed from there.}
\end{figure}

\begin{figure}[htb]  
\centering
\includegraphics[width=\columnwidth]{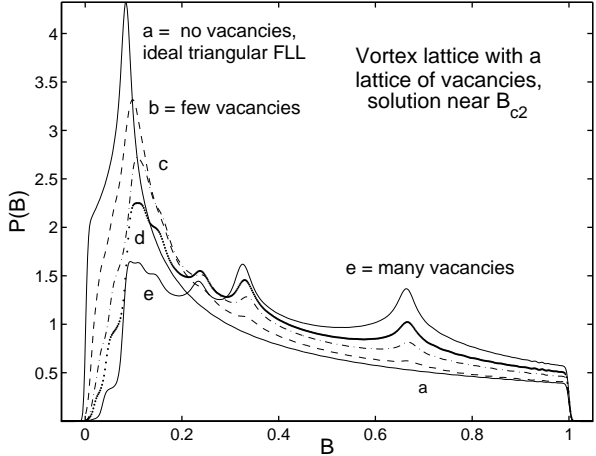}
\caption{$P(B)$ Field density $P(B)$ of a vortex lattice
with various vacancy concentrations $1/N^2$:
a) no vacancy, b) $N=9$, c) $N=6$, d) $N=4$, e) $N=3$.
GL solution near $B_{c2}$.}
\end{figure}

\section{Vortex lattice containing vacancies}

  As an example for structural defects I consider a vortex
lattice (spacing $a$) with a super lattice of vacancies
with spacing $Na$, $N=2$, 3, 4, $\dots$. This problem was
solved in \cite{2} both within London theory (at $b \ll 1$)
and GL theory near $b=1$ where
  \begin{eqnarray}  
  B({\bf r}) = \bar B + [\, \langle \omega \rangle -
  \omega({\bf r}) \,]\, B_{c2}/ (2 \kappa^2)
  \end{eqnarray}
holds, i.e., the shapes of $\omega(x,y)$ and $B(x,y)$ are the
same. For this vacancy lattice near $b=1$ one has
  \begin{equation}  
  \omega({\bf r}) = c_1 \,{ \omega_A({\bf r}) \over
  \omega_A({\bf r}/N) }   \prod_\nu {
  \omega_A[ ({\bf r-R}_\nu -{\bf s}_\nu) /N ]    \over
  \omega_A[ ({\bf r-R}_\nu             ) /N ] } ,
  \end{equation}
where $\omega_A(x,y)$ is the Abrikosov $B_{c2}$ solution
given below Eq.~(12), $c_1$ is a normalization constant,
the product is over all vortex positions
${\bf R}_\nu= {\bf R}_{mn}$ within the super cell,
and the vortex displacements ${\bf s}_\nu $
are chosen such as to minimize the free energy and
the Abrikosov parameter
$\beta=\langle\omega^2 \rangle/\langle \omega\rangle^2 >1$.
This relaxation of the vortex positions around the vacancy
yields an $\omega(x,y)$ with nearly constant spatial
amplitude, i.e, the maximum $\omega(0,0)$ at the vacancy
position has about the same height as all the maxima
of $\omega$ between the vortex positions.

Figure~6 shows the contours of $\omega$ (and thus of $B$)
for a vortex lattice with one vacancy (limit $N\to \infty$)
at the origin and with central symmetric displacements
${\bf s}_\nu = - {\bf R}_\nu
 [\sqrt{3} a^2 / (4\pi R_\nu^2) + 0.068 a^4/ R_\nu^4 ]$.
The field density $P(B)$ of vortex lattices with various
vacancy concentrations $1/N^2$ is shown in Fig.~7.
The new peaks indicate that new saddle points and
minima of $B(x,y)$ (i.e., maxima of $\omega$)
appear near the vacancy, as seen also in Fig.~6.

\begin{figure}[htb]   
\centering
\includegraphics[width=\columnwidth]{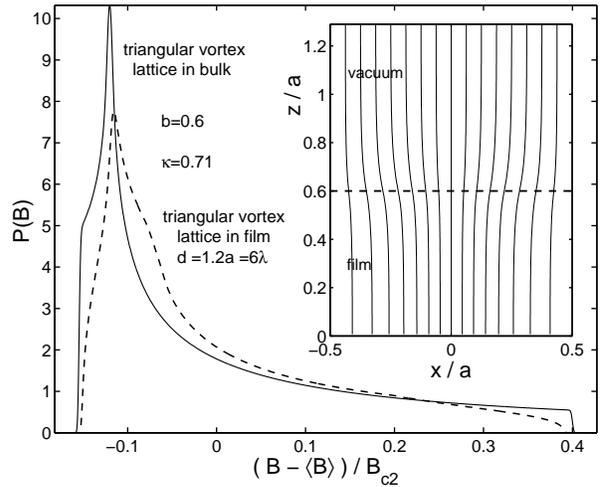}
\caption{The field density $P(B)$ of the triangular vortex lattice
in a superconducting film of thickness $d = 1.2 a = 6 \lambda$
in a perpendicular field $B_a = \bar B$ for $\kappa = 0.707$,
$b=\bar B/B_{c2}=0.6$, from GL theory. Shown are
$P(B)$ for bulk (2D, solid line) and for film (3D, dashed
line). The inset shows the magnetic field lines in and near
the film; the dashed line marks the film surface
$z =d/2 =0.6a$, $a$ = vortex spacing.}
\end{figure}

\section{3D vortex lattice in films}

The Fourier method of Sec.~2 may be generalized to
superconducting films of arbitrary thickness $d$
($|z| \le d/2$) containing
a vortex lattice that is periodic in the $(x,y)$
plane. For this one has to add to $F$, Eq.~(5), the energy
$F_{\rm stray}$ of the magnetic stray field outside the
film and has to use $z$ dependent trial functions,
e.g., (6)-(8) with Fourier coefficients depending on
3D vectors ${\bf K} = (K_x, K_y, K_z)$ \cite{15}. When the
applied field $B_a$ is along $z$ (perpendicular to the
film plane) then the short vortex lines are straight and
along $z$. The inset of Fig.~8 shows how the magnetic
field lines of the vortices become less dense when they
approach the film surface, they ``mushroom'' and
go over smoothly into the stray field.

Though in the
depicted example ($b=0.6$, $\kappa=0.707$,
$d=1.2 a \approx 6\lambda \approx 4 \xi$) this widening
of the field lines is only weak, it still has a strong
effect on the field density $P(B)$ shown as dashed line,
namely, the jumps of $B_{\rm min}$ and $B_{\rm max}$
and the saddle-point peak are smeared and shifted
towards $\bar B$, and $P(B)$ is increased (has a hump)
between $B=B_{\rm sad}$ and $\bar B$. These features
are expected since for very thin films with
$d < \lambda$ the field amplitude is strongly
reduced inside the film and $P(B)$ narrows to a line
positioned at $B = \bar B$. This $P(B)$ should be
observable by $\mu$SR in specimens
composed of many thin layers separated by a distance
$\ge a/4$ where the stray-field modulation amplitude
$\propto \exp[ -2\pi (|z|-d/2)/a]$ has almost vanished.

  In such infinitely extended films one has $\bar B = B_a$
since all field lines have to pass the film.
The magnetization $M$ of the film, therefore, cannot be
calculated as a difference of fields,
but one has to take the derivative of the total free energy,
$M = -\partial (F + F_{\rm stray}/d) /\partial \bar B$.
A more elegant method calculates $M$ by Doria's
virial theorem directly from the GL solution for the film,
with no need to take an energy derivative \cite{16}.

\section{Random perturbations}

In real superconductors randomly positioned weak pinning
centers, or random pinning forces, may lead to more or
less random small displacements of the vortices from their
ideal lattice positions. As shown in \cite{11}, in bulk
superconductors this leads approximately
to a convolution of the ideal-lattice $P(B)$ with a
Gaussian. Computer simulations of this problem based on
London theory (i.e., pairwise interacting vortex lines
and linear superposition of vortex fields) are presented
in \cite{11}. Interestingly, while disorder of a 2D vortex
lattice {\bf broadens} $P(B)$ and its singularities,
disorder in the 3D point-vortex (or pancake-vortex)
lattice occurring in layered high-$T_c$
superconductors \cite{6,17} typically will lead
to {\bf narrowing} of $P(B)$ \cite{18}
since the vortex lines (pancake stacks) become wider.
A further effect that contributes to the broadening of
$P(B)$ is the (quantum) diffusion of muons after they
have stopped, e.g., in ultrapure Nb \cite{19,20,21}.

 Improved pinning simulations using GL theory and
considering also {\bf thermal fluctuations} of vortices
are desirable, as well as microscopic calculations going
beyond the GL approach. From BCS-Gor'kov theory it is
shown in \cite{11,19} that for pure Nb near $B_{c2}$
the $P(B)$ depends on temperature $T$, and at
$T \ll T_c$ it looks quite different from the GL result
valid near the critical temperature $T_c$ since then
$B(x,y)$ has sharp conical maxima and minima, and two
saddle points with three-fold symmetry yielding
an infinity of the form
$P(B) \propto |B - B_{\rm sad}|^{-1/3}$.




\begin{thebibliography}{00}

\bibitem{1} A.~A.~Abrikosov, Zh.~Exp.~Teor.~Fiz.~{\bf 32}, 1442,
            1957 (Sov.\ Phys.-JETP {\bf 5}, 1174, 1957).
\bibitem{2} E.~H.~Brandt, phys.~stat.~sol. {\bf 35}, 1027 (1969);
            {\bf 36}, 371 (1969); {\bf 36}, 381 (1969);
            {\bf 36}, 393 (1969).
\bibitem{3} E.~H.~Brandt, J.~Low.~Temp.~Phys.~{\bf 26}, 709 (1977);
            {\bf 26}, 735 (1977); {\bf 28}, 263 (1977);
            {\bf 28}, 291 (1977).
\bibitem{4} E.~Zeldov et al., Nature {\bf 375}, 373 (1995).
\bibitem{5} A.~M.~Campbell and J.~E.~Evetts, Adv.~Phys.~{\bf 21},
            199 (1972).
\bibitem{6} G.~Blatter, M.V.~Feigel'man, V.B.~Geshkenbein,
            A.I.\ Larkin, V.M.\ Vinokur, Rev.\ Mod.\ Phys. {\bf 66}
            (1994) 1125.
\bibitem{7} K.~Shibata, T.~Nishizaki, T.~Sasaki, N.\ Kobayashi,
            Phys.~Rev.~B {\bf 66}, 214518 (2002).
\bibitem{8} G.~P.~Mikitik and E.~H.~Brandt, Phys.\ Rev.\ B {\bf 64}
            184514 (2001); dito {\bf 68}, 054509 (2003)
\bibitem{9} I.~L.~Landau and H.~Keller, Phys.~C {\bf 466}, 131
            (2007).
\bibitem{10} E.~H.~Brandt, Phys.~Rev.~B {\bf 64}, 024505 (2001).

\bibitem{11} E.~H.~Brandt, J.~Low Temp.~Phys.~{\bf 73}, 355 (1988).

\bibitem{12} E.~H.~Brandt, Phys.~Rev.~Lett. {\bf 78}, 2208 (1997).

\bibitem{13} E.~H.~Brandt, Phys.~Rev.~B {\bf 68}, 054506 (2003).

\bibitem{14} E.~H.~Brandt, Rep.~Prog.~Phys.~{\bf58}, 1465-1594 (2003).

\bibitem{15} E.~H.~Brandt, Phys.~Rev.~B {\bf 71}, 014521 (2005).

\bibitem{16} M.~M.~Doria, E.~H.~Brandt, and F.~M.~Peeters,
             Phys.~Rev.~B (in print).
\bibitem{17} J.~R.~Clem, Phys.~Rev.~B {\bf 43}, 7837 (1991).
\bibitem{18} E.~H.~Brandt, Phys.~Rev.~Lett. {\bf 66}, 3213 (1991).

\bibitem{19} E.~H.~Brandt and A.~Seeger, Adv.~Physics {\bf 35},
             189 (1986).
\bibitem{20} A.~Schwarz et al., Hyperfine Interact.~{\bf 31},
             247 (1987).
\bibitem{21} D.~Herlach et al., Hyperfine Interact.~{\bf 63},
             41 (1990).





\end{thebibliography}
\end{document}